\documentclass[twocolumn,final,aps,prd,showpacs,amsmath,amssymb,amsfonts,floatfix,longbibliography]{revtex4-1}
\usepackage{graphicx}
\usepackage{multirow}
\usepackage{epsfig}
\usepackage{amsmath}

\def\be{\begin{equation}}
\def\ee{\end{equation}}

\usepackage{color}
\usepackage[colorlinks,breaklinks,bookmarks=true,citecolor=blue,linkcolor=red,urlcolor=blue]{hyperref}

\begin{document}

\title{Finite-temperature phase diagram of (111) nickelate bilayers}

\author{Oleg Janson}
\email{olegjanson@gmail.com}
\affiliation{Institut f\"{u}r Festk\"{o}rperphysik, TU Wien, Wiedner
Hauptstra{\ss}e 8-10, 1040 Vienna, Austria}

\author{Karsten Held}
\affiliation{Institut f\"{u}r Festk\"{o}rperphysik, TU Wien, Wiedner
Hauptstra{\ss}e 8-10, 1040 Vienna, Austria}
\date{\today}

\begin{abstract}
We report a density functional theory plus dynamical mean field theory
(DFT+DMFT) study of an oxide heterostructure of LaNiO$_3$ (LNO) bilayers in
$[$111$]$ direction interleaved with four atomic monolayers of LaAlO$_3$.
DFT+$U$ optimizations yield two stable solutions: a uniform structure with
equivalent NiO$_6$ octahedra, as well as a bond-disproportionated (BD)
structure featuring a breathing distortion.  For both structures, we construct
the low-energy models describing the Ni $e_g$ states by means of Wannier
projections  supplemented by the Kanamori interaction, and solve them by DMFT.
Using the continuous-time quantum Monte Carlo algorithm in the hybridization
expansion, we study the temperature range between 145 and 450\,K.  For the
uniform and the BD structure, we find similar phase diagrams that comprise four
phases: a ferromagnetic metal (FM), a paramagnetic metal (PM), an
antiferro-orbitally-ordered insulator (AOI), as well as a paramagnetic
insulator (PI).  By calculating momentum-resolved spectral functions on a torus
and a cylinder, we demonstrate that the FM phase is not a Dirac metal, while
both insulating phases are topologically trivial.  By a comparison with
available experimental data and model DMFT studies for the two-orbital Hubbard
model, we suggest that LNO bilayers are in the AOI phase at room temperature.
\end{abstract}


\maketitle

\section{Introduction}
Transition metal oxides (TMO) exhibit a plethora of fascinating physical
behaviors, such as metal-insulator transitions~\cite{imada98},
multiferroicity~\cite{cheong07,tokura10,tokura14}, colossal
magnetoresistance~\cite{ramirez97, tokura06}, and high-temperature
superconductivity~\cite{pickett89,lee06}.  The ongoing progress in fabricating
high-quality TMO thin films led to the emergence of a new class of artificial
materials: oxide heterostructures.  Their properties are often markedly
different from their TMO constituents.  An archetypical example are
heterostructures of two band insulators, SrTiO$_3$ (STO) and LaAlO$_3$ (LAO):
if the latter component reaches the critical thickness of four atomic
monolayers, a two-dimensional (2D) electron gas emerges at the
interface~\cite{ohtomo04}.  Even more striking effects are observed for TMO
with a partly filled $d$ shell. Here, electronic correlations become essential.
For example, the emergence of a ferromagnetic metal at LaMnO$_3$/SrMnO$_3$
interfaces was reported~\cite{bhattacharya08}, where both constituents are bulk
antiferromagnetic insulators; or the polar field and the Mott insulating gap of
STO/LaVO$_3$ can be employed as a solar cell~\cite{assmann13, wu15}.

Although the variety of TMO gives rise to a very large number of possible
binary combinations, the count of studied oxide heterostructures grows slowly.
Following the pioneering works of Hwang and Ohtomo~\cite{ohtomo02,ohtomo04},
the research has been largely focused on superlattices whose constituents have
a (possibly distorted) perovskite structure in the bulk, and the direction of
growth was typically chosen to be along the (pseudo)cubic $[$001$]$ direction.
Against this backdrop, Xiao \emph{et~al.}~\cite{xiao11} argued that bilayers
grown along the trigonal axis, i.e.\ in the $[$111$]$ direction, form a
honeycomb lattice, with an excellent potential to create correlated analogs of
graphene~\cite{xiao11}.  By employing the tight-binding (TB) approximation,
they studied different fillings of the correlated $d$ shell in the presence of
the spin-orbit coupling (SOC), and demonstrated that such (111) bilayers can
host various topologically nontrivial phases.  Also Haldane's~\cite{haldane88}
quantum anomalous Hall state can be realized in (111) bilayers of SrRuO$_3$ on
STO~\cite{si17}.

\begin{figure}[tb]
\includegraphics[width=8.6cm]{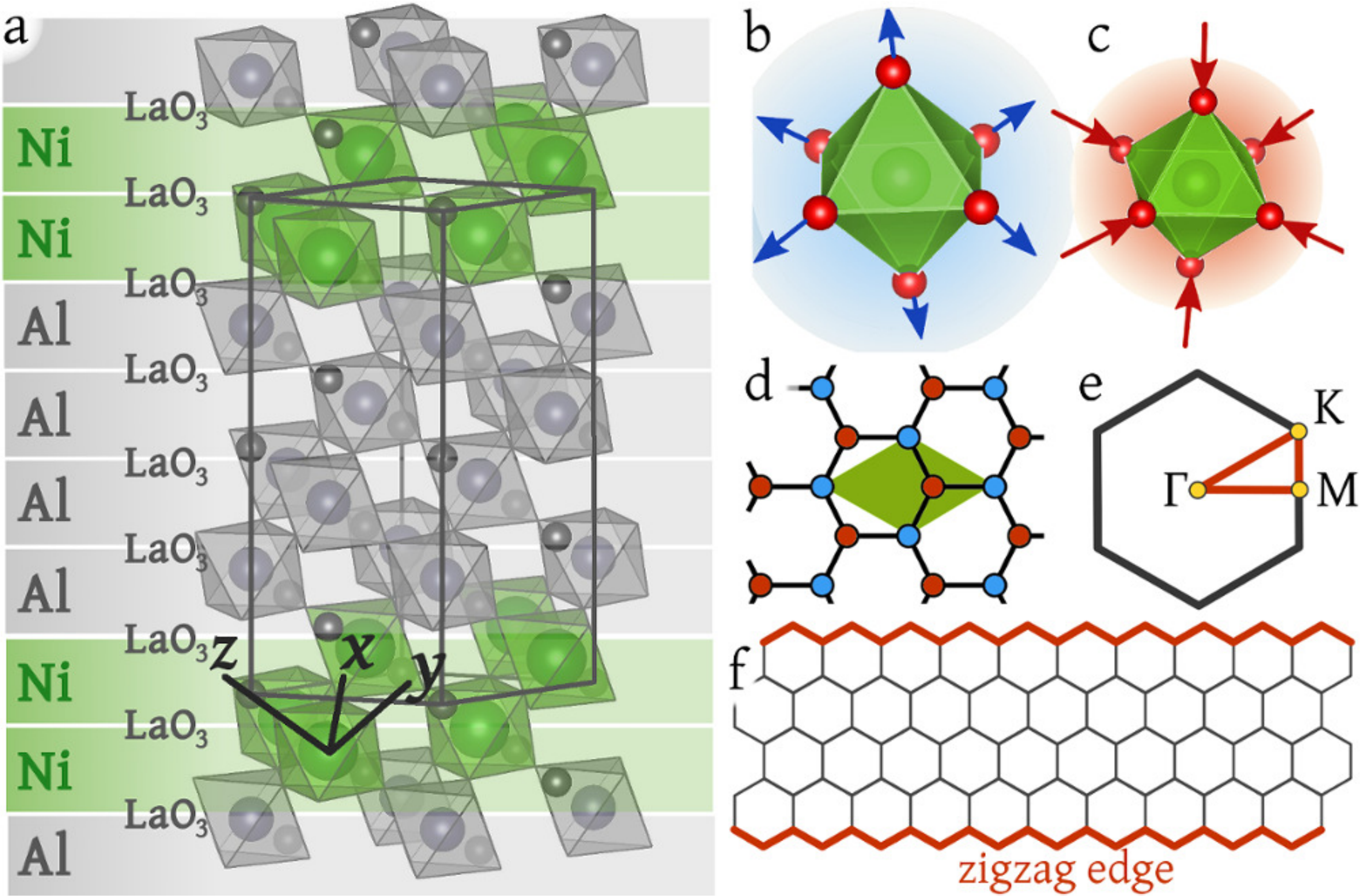}
\caption{\label{fig:str}(Color online) (a) In the trigonal unit cell used for
DFT+DMFT calculations, (111) nickelate bilayers are separated by four LaAlO$_3$
layers. For NiO$_6$ octahedra, the local coordinate axes $x,y,z$ are indicated.
Bond disproportionation gives rise to a breathing distortion: stretched (b) and
squeezed (c) NiO$_6$ octahedra alternate in the lattice (d).  (e)
High-symmetry points of the Brillouin zone (BZ). (f) Zigzag edges of the
honeycomb lattice used for the edge state calculations.}
\end{figure}

The ensuing numerical studies extended the TB analysis by including electronic
interaction effects on a mean-field level~\cite{yang11,ruegg11}, and
identified LaNiO$_3$ bilayers (2LNO) in an LAO matrix as a promising candidate
for the realization of topological states~\cite{yang11,ruegg12}.  In the
simplest ionic approximation, Ni$^{3+}$ has the $d^7$ electronic configuration,
whereby six electrons fully occupy the low-lying $t_{2g}$ states and render
them inactive.  Hence, all change, orbital, and spin degrees of freedom in the
2LNO/$n$LAO superlattices (Fig.~\ref{fig:str}, a) pertain to the single
electron in the $e_g$ orbitals that remain degenerate in the trigonal
symmetry~\cite{xiao11}.  This resilient degeneracy, in contrast to
(001) superlattices~\cite{hansmann09}, gives room for spontaneous ordering of
complex orbitals and thus topologically nontrivial states emerge despite the
small SOC~\cite{ruegg11}.  While first analyses involving realistic
tight-binding Hamiltonians evaluated by means of density functional theory
(DFT)~\cite{ruegg12} as well as DFT+$U$ calculations~\cite{ruegg12,ruegg13}
suggested the stability of a Dirac semimetal state, later DFT+$U$ studies
established a key role of a breathing distortion of NiO$_6$ octahedra
(Fig.~\ref{fig:str} b, c) which opens a gap and competes with the topological
states~\cite{ruegg13, doennig14}.  The breathing distortion is accompanied by a
polarization, rendering (111) LNO bilayers a prospective multiferroic with a
sizable spin polarization~\cite{doennig14}.

On the experimental side, transport measurements indicated a
semiconducting behavior, with the gap showing a sizable dependence on the
thickness of the LAO layer~\cite{middey12}.  Recently, both the activated
behavior and the sensitivity of the gap size were corroborated by an
independent study, which reported gaps between 17 and 162\,meV for different
LAO thicknesses~\cite{wei16}.  The nature of the gap, in particular whether it
is topological or related to a breathing distortion, remains an open question.  

Here, we employ a combination of DFT and dynamical mean-field theory
(DMFT)~\cite{anisimov97,lichtenstein98,kotliar06,held07,janson18} to explore
the phase diagram of 2LNO/4LAO heterostructures (Fig.~\ref{fig:str}, a).  By
performing detailed structural relaxations, we demonstrate that the presence of
a breathing distortion can be neither proved, nor disproved: both structure
types feature the same DFT+$U$ total energies within the error bars.  Hence, we
carry out DFT+DMFT calculations for the uniform as well as the
bond-disproportionated structure.  Both show, despite the symmetry breaking
associated with the distortion, actually quite similar physics.

This paper is organized as follows.  The methods employed, including the
structure optimization, DMFT, and the evaluation of spectral functions, are
described in Sec.~\ref{sec:method}. DFT+DMFT results for the uniform and the
bond-disproportionated structures are presented in Sec.~\ref{sec:results}.  A
discussion of the topological properties as well as comparison of our numerical
results with the available experimental data are given in Sec.~\ref{sec:disc}.
We conclude our paper and provide a brief outlook in Sec.~\ref{sec:summary}.

\section{\label{sec:method}Method}
\subsection{Optimization of the crystal structure}
DFT+DMFT results generally depend on the structural input from DFT
calculations. Hence, accurate information on the crystal structure is of
crucial importance.  This is particularly challenging for superlattices that
are not amenable to standard x-ray or neutron diffraction measurements.  A
common approach is to evaluate the structural input computationally, by
allowing for a relaxation of the atomic coordinates, but keeping the lattice
constants fixed to that of the substrate and minimizing the total
energy~\cite{janson18}.  In contrast, for correlated materials, the
underestimation of electronic correlations can have drastic impact on the
crystal structure.  A prominent example is KCuF$_3$, where orbital ordering can
give rise to a distortion of the lattice, known as the cooperative Jahn-Teller
effect.  This can be assisted by lattice effects, which may also be a driving
force.  While conventional DFT functionals yield a spurious undistorted
structure, DFT+$U$ captures the underlying physics and reproduce the
experimentally observed distortion~\cite{liechtenstein95}.

A distinct trait of bulk nickelates is their tendency towards bond
disproportionation, i.e.\ developing a breathing distortion of NiO$_6$
octahedra~(e.g.,~\cite{alonso99, medarde08, garcia-munoz09}).  Although it is
not the case for bulk LaNiO$_3$, the compressive strain exerted by the
LaAlO$_3$ substrate can stabilize the respective distortion.  Similarly to
KCuF$_3$, this physics is not captured by DFT, and conventional functionals
disfavor such a bond disproportionation~\cite{ruegg13}.  Hence, structural
optimizations of nickelate superlattices performed using a conventional DFT
functional can possibly lead to spurious results, and correlations have to be
accounted for in the course of a structural optimization.  The optimal solution
would be a self-consistent DFT+DMFT scheme with an atomic force calculation at
every step, but such calculations require enormous computational efforts and
remain unfeasible for multisite and multi-orbital systems such as nickelate
heterostructures.  Therefore, in this work, we restrict ourselves to DFT+$U$
structural optimizations that generally capture the structural details in the
rare earth nickelates~\cite{hampel17}.

We employ the generalized gradient approximation (GGA) +$U$ functional with $U$
in the range 4.0 to 6.0\,eV and $J$\,=\,1.0\,eV as implemented in
\textsc{vasp-5.3}~\cite{vasp, *vasp_2}.  Ionic relaxations are performed until
all forces are below 0.005\,eV/\r{A}.  The in-plane unit cell parameters are
fixed to that of bulk LAO. To optimize the $c$ parameter, we construct cells
with different $c$ and subsequently optimize the atomic coordinates.  In this
way, we find that $c$\,=\,13.30\,\r{A} yields the lowest total energy
independent of the $U$ value.  Next, we consider two trial structures as a
starting point --- a uniform structure and a structure with a breathing
distortion (BD), see Fig.~\ref{fig:str} b,c --- and relax the atomic
coordinates by keeping the unit cell parameters fixed.  Despite the general
trend that the uniform structure has a lower energy for smaller $U_d$, the
energy differences are of the order of several K per cell, i.e.\ on a par with
the accuracy of DFT total energies.  We conclude that the elastic energy due to
BD and the concomitant change in the electric potential are well-balanced, so
that DFT+$U$ calculations cannot provide an unambiguous answer, which of the
two structure types is realized in 2LNO/4LAO.  We therefore perform DFT+DMFT
calculations for both, the uniform as well as the BD structure (Table~\ref{tab:poly}).

\begin{table}[h]
\caption{\label{tab:poly} Comparison of the relevant structural parameters in
the uniform as well as the bond-disproportionated (111) 2LNO/4LAO structures
optimized in the GGA+$U$ with the experimental structure of bulk
LaNiO$_3$~\cite{garcia-munoz92}.  Atomic coordinates of the uniform and the
bond-disproportionated structure are provided in the Appendix.
}
\begin{ruledtabular}
\begin{tabular}{rrrr}
structure & $\langle$Ni--O$\rangle$, \r{A} & $V_{\text{NiO$_6$}}$, \r{A}$^3$ & $\measuredangle_{\text{Ni--O--Ni}}$, $^{\circ}$\\ \hline
uniform                 & 1.938 & 9.70 & 166.05 \\
BD                      & 1.952\,/\,1.925 & 9.91\,/\,9.52 & 166.04 \\
bulk LaNiO$_3$ (298\,K) & 1.935 & 9.65 & 165.22 \\
\end{tabular}
\end{ruledtabular}
\end{table}

\subsection{DFT+DMFT}
Subsequently, self-consistent DFT calculations for both optimized structures
(uniform and BD) were performed using \textsc{wien2k}~\cite{wien2k}. Both GGA
band structures (not shown) feature a well-separated manifold crossing
the Fermi level, the respective bands are formed by the antibonding combination
of Ni $e_g$ and O $p$ states. Since we have two Ni atoms per cell (see
Fig.~\ref{fig:str}, a, d) and two $e_g$ orbitals per Ni, the total number of
bands in the manifold is four.  For these four bands, we construct the
maximally localized Wannier functions (WF) using
\textsc{wannier90}~\cite{wannier90} via the \textsc{wien2wannier}
interface~\cite{wien2wannier}.  Finally, we calculate $H(k)$ by Fourier
transforming the Wannier Hamiltonian on a 48$\times$48$\times$1 $k$-mesh. 

DMFT calculations were performed using the continuous-time quantum Monte Carlo (CT-QMC)
in the hybridization expansion (CT-HYB)~\cite{werner06} as implemented in
\textsc{w2dynamics}~\cite{parragh12, w2dynamics}. We used the rotationally
invariant Kanamori interaction $U'\!=\!U\!-\!2J$, which, in addition to the
density-density interaction, accounts also for the spin flip and pair hopping
terms.  By fixing the Hund's exchange $J$ to 0.75\,eV~\footnote{Note that SU(2)
symmetric interactions require a smaller $J$ compared to the Slater
parametrization used in DFT+$U$: $U_\text{Kanamori} = U_\text{Slater} +
\frac{8}{7}J_\text{Slater}$ and $J_\text{Kanamori} =
\frac{5}{7}J_\text{Slater}$. See Supplementary Note\,1 in \cite{hausoel17} for
details.}, we varied the inter-orbital Coulomb repulsion $U'$ from 2 to
5\,eV~\footnote{The $U_d$ value in DMFT calculation is considerably smaller
than $U_d$ in DFT+$U$ due to the different spatial extent of the orbitals to
which they are applied: DMFT operates on Wannier orbitals, having substantial
Ni and O contributions, while in DFT+$U$, repulsion pertains to the spatially
confined atomic $d$-orbital of Ni.} and scanned the temperature range between
145 and 450\,K ($80\geq\beta\geq25$\,eV$^{-1}$). At every DMFT step, two
independent two-orbital impurity problems (for each Ni atom) in the unit cell
were solved.  State-of-the-art DFT+DMFT calculations involve full
charge-self-consistency, which plays an important role in heterostructures with
a sizable electron transfer~\cite{lechermann13, *lechermann15}.  Here, we
employed the non-charge-self-consistent scheme, i.e.\ the chemical potential
was adjusted to have two electrons per unit cell in each DMFT iteration.  We
however found that the resulting per-atom occupations in DMFT were similar to
the GGA ones. The absence of appreciable charge transfer between the Ni sites
\emph{a posteriori} justifies the usage of a non-charge-self-consistent
DFT+DMFT~(see the discussion in Ref.~\onlinecite{bhandary16}).

The quasiparticle renormalization is estimated from the slope of the imaginary
part of the self-energy $\text{Im}{\Sigma(i\omega_n)}$ which depends on
$\omega_n$ linearly at low Matsubara frequencies:

\begin{equation}
\label{eq:Z}
Z\simeq\left(1-\frac{\partial\text{Im}{\Sigma(i\omega_n)}}{\partial\omega_n}\right)^{-1}.
\end{equation}

For each DMFT calculation, we considered only those
$\text{Im}{\Sigma(i\omega_n)}$ that still lie on a straight line according to a
$\chi^2$ fit.

\subsection{Spectral functions}
For selected values of $\beta$ and $U$, we present spectral functions.  For
these we used self-energies $\Sigma(i\omega)$ computed with the worm
algorithm~\cite{gunacker15}, and analytically continued them to the real axis using
\textsc{Maxent}~\cite{levy17} which employs the maximum entropy
method~\cite{jarrell96}.  The resulting self-energies $\mathbf{\Sigma}(\omega)$
are used to calculate the interacting Green's function on the real frequency
axis:

\begin{equation}
\label{eq:Gw}
\textbf{G}^{-1}(\vec{k},\omega) = \left(\omega\!+\!i\delta\!+\mu\right)\mathbf{I}\!-\!\mathbf{H}(\vec{k})\!-\!\mathbf{\Sigma}(\omega)\!-\!\mathbf{\Sigma}_{\text{dc}},
\end{equation}
where matrices in terms of orbitals and sites of the cell are denoted bold,
$\mathbf{\Sigma}_{\text{dc}}$ is the double-counting correction in the fully
localized limit~\cite{anisimov93} and $\mathbf{I}$ the identity matrix.  The
$\vec{k}$-resolved and $\vec{k}$-integrated spectral functions can be obtained as

\begin{equation}
\label{eq:Awk}
A(\vec{k},\omega) = -\frac{1}{\pi}\left(\frac{1}{m}\right)\text{Tr}\left[\text{Im}\textbf{G}(\vec{k},\omega)\right]
\end{equation}

and 

\begin{equation}
\label{eq:Aw}
A(\omega) = \left(\frac{1}{N_{\vec{k}}}\right)\sum_{\vec{k} \in \text{BZ}}{A(\vec{k},\omega)},
\end{equation}

respectively, where $m$ is the dimension of the $\textbf{H}(\vec{k})$ matrix.
These spectral functions are based on $H(\vec{k})$ with the periodic boundary
conditions along both in-plane directions, i.e.\ the Hamiltonian is defined on
a torus.  To address the edge states, we resort to mixed boundary conditions of
a cylinder, which is periodic along $x$ and open along $y$~\footnote{ We note
that the honeycomb lattice allows for two inequivalent terminations: the zigzag
edge and the armchair edge.  All calculations in this study are performed for
the zigzag edge.}.  The respective Hamiltonian $\textbf{H}(k_x)$ is now a
$n_ym\times{}n_ym$ matrix, where $n_y$ is the number of unit cells along the
open direction, and the Green's function is

\begin{equation}
\label{eq:Gwedge}
\textbf{G}^{-1}(k_x,\omega) = \left(\omega\!+\!i\delta\!+\mu\right)\mathbf{I}-\!\mathbf{H}(k_x)\!-\!\mathbf{\Sigma}(\omega)\!-\!\mathbf{\Sigma}_{\text{dc}}.
\end{equation}

The respective spectral function is

\begin{equation}
\label{eq:Awkx}
A(k_x,\omega) = -\frac{1}{\pi}\left(\frac{1}{n_ym}\right)\text{Tr}\left\{\text{Im}\left[\textbf{G}(k_x,\omega)\right]\right\}.
\end{equation}

Since we are primarily interested in the edge states, we also explicitly
calculate their contribution to the spectral weight as:

\begin{equation}
\label{eq:Awkx_edge}
A^{\text{edge}}(k_x,\omega) = -\frac{1}{\pi}\left(\frac{1}{2m}\right)\text{Tr}\left\{\text{Im}\left[\textbf{G}_{\text{TT}}(k_x,\omega) +\\ \textbf{G}_{\text{BB}}(k_x,\omega)\right]\right\},
\end{equation}
where $\textbf{G}_{\text{TT}}$ ($\textbf{G}_{\text{BB}}$) denotes the Green's
function projected onto the top (bottom) cell.

\subsection{Choice of the model}
Before turning to the DMFT results, we address a controversially discussed
issue of the minimal model for nickelates.  The hybridization of Ni $e_g$
states and the $\sigma$-bonded O $p$ states gives rise to molecular-like
$dp_{\sigma}$ orbitals. In a unit cell of $n$ Ni atoms, the antibonding states
form an isolated $\frac14$-filled manifold of $2n$ bands at the Fermi energy.
For low-energy excitations, it is seemingly natural to restrict the analysis to
these states and use the respective antibonding $dp_{\sigma}$ orbitals as a
basis in real space.  This minimal two-orbital model, known as the $d$-only
model, has been employed in early DMFT studies~\cite{hansmann09, hansmann10}.

On the other hand, the high oxidation state of Ni$^{3+}$ can lead to a very
small, possibly even negative charge transfer gap.  In this case, also the
low-energy physics will be largely affected by charge transfer processes
between $d$ and $p$ states.  Indeed, DMFT calculations for such $d+p$ models
yielded qualitatively different results~\cite{han11}, mainly because the
$e_g^2$ oxygen ligand hole (L) configuration resulting from the negative charge
transfer forms a spin $S$\,=\,1 on the Ni sites~\cite{parragh13}.  It has been
further suggested that every second Ni site forms a spin singlet with two
ligand holes~\cite{park12, green16, haule17}, leaving only localized  $S=1$
states on the other half of the Ni sites. One should carefully note however
that whether one has a negative charge transfer ($d^8$) or not ($d^7$) very
sensitively depends on the relative position of oxygen and Ni $e_g$ states. In
DFT, the oxygen bands are too close to the Fermi level, which would favor the
negative charge transfer $d^8L$  picture.  On top of this, the DFT+DMFT double
counting and possible inclusion of the $d$-$p$ interaction make a theoretical
prediction unreliable. Hence, in our view, this question has to be answered by
experiment eventually.  In this respect, there are indications of a $d^8L$
configuration from x-ray absorption spectroscopy for smaller rare earth cations
such as NdNiO$_3$~\cite{bisogni16}, but not for bulk LaNiO$_3$: very recent
single-crystal experiments~\cite{guo18} yield the ordered magnetic moment of
$~\sim$0.3\,$\mu_{\text{B}}$, which is far too low for $S$\,=\,1.

In fact a BD scenario can be realized also in a $d$-only model as has been
acknowledged long ago~\cite{mazin07}. Recent DFT+DMFT calculations by
Subedi~\emph{et al.} showed that the BD phase sets in if $(U_d-3J_d)$ is
smaller than the difference between the on-site energies of the $e_g$
orbitals~\cite{subedi15}, which in our case is zero (degenerate $e_g$
orbitals).  This result demonstrates that the emergence of the BD phase, and
hence, the nature of the metal-insulator transition in bulk nickelates are
reproduced by a $d$-only model, albeit with a strongly reduced Coulomb
interaction $U_d$. 

In view of this and the unclear experimental situation, we restrict ourselves
to the $d$-only model. It features a considerably smaller number of free (and
prospectively very sensitive) parameters; and because the effective Coulomb
repulsion in the $d$-model can be strongly reduced~\cite{subedi15}, we scan a
broad range of $U_d$.

\section{\label{sec:results}DFT+DMFT results}
\subsection{Uniform structure}
\begin{figure}[tb]
\includegraphics[width=8.6cm]{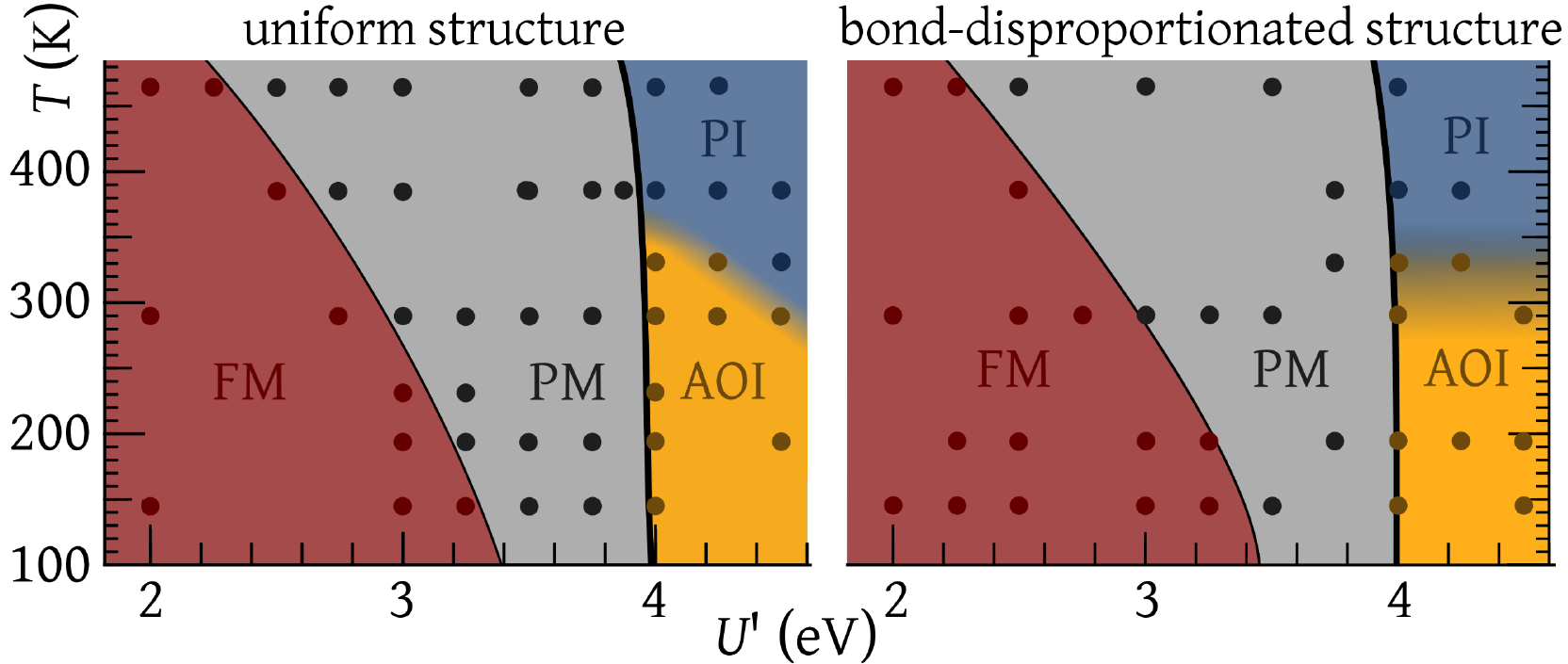}
\caption{\label{fig:phasediag}(Color online) DFT+DMFT phase diagram for the
uniform structure (left) and the bond disproportionated structure (right).  FM,
PM, PI, and AOI stand for ferromagnetic metal, paramagnetic metal, paramagnetic
metal, and antiferro-orbitally-ordered insulator.  Every thick point denotes a
separate DMFT calculation.}
\end{figure}

\begin{figure}[tb]
\includegraphics[width=8.6cm]{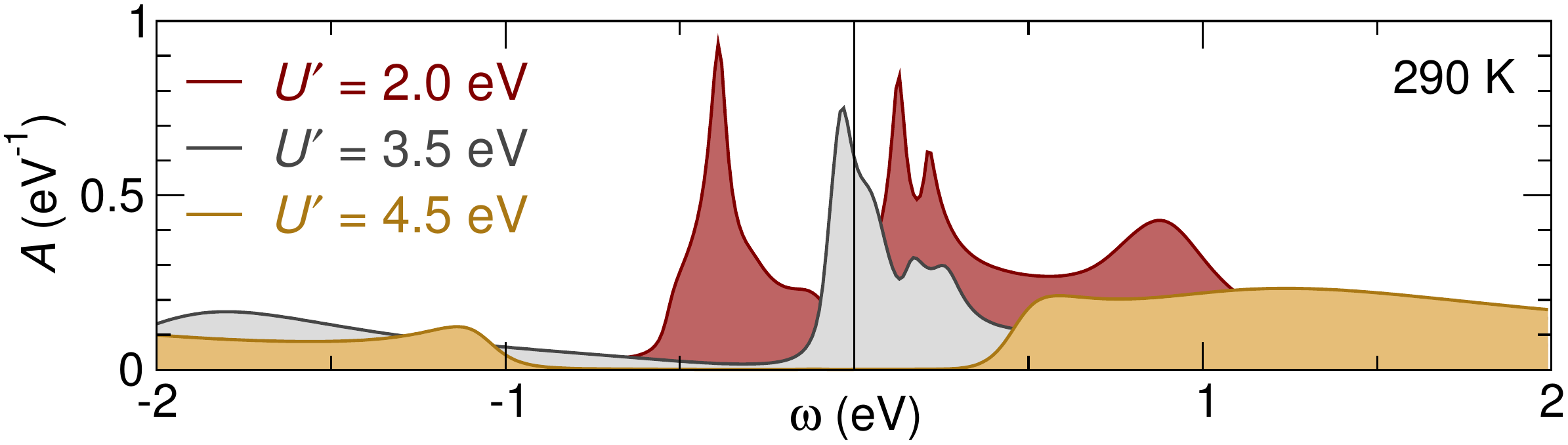}
\caption{\label{fig:aw}(Color online) Spectral functions $A(\omega)$
$[$Eq.~(\ref{eq:Aw}$)]$ of the uniform structure calculated with DMFT at room
temperature ($T$\,=\,290\,K) for the FM phase ($U'$\,=\,2\,eV), the PM phase
($U'$\,=\,3.5\,eV) and the AOI phase ($U'$\,=\,4.5\,eV)} \end{figure}

\begin{figure}[tb]
\includegraphics[width=8.6cm]{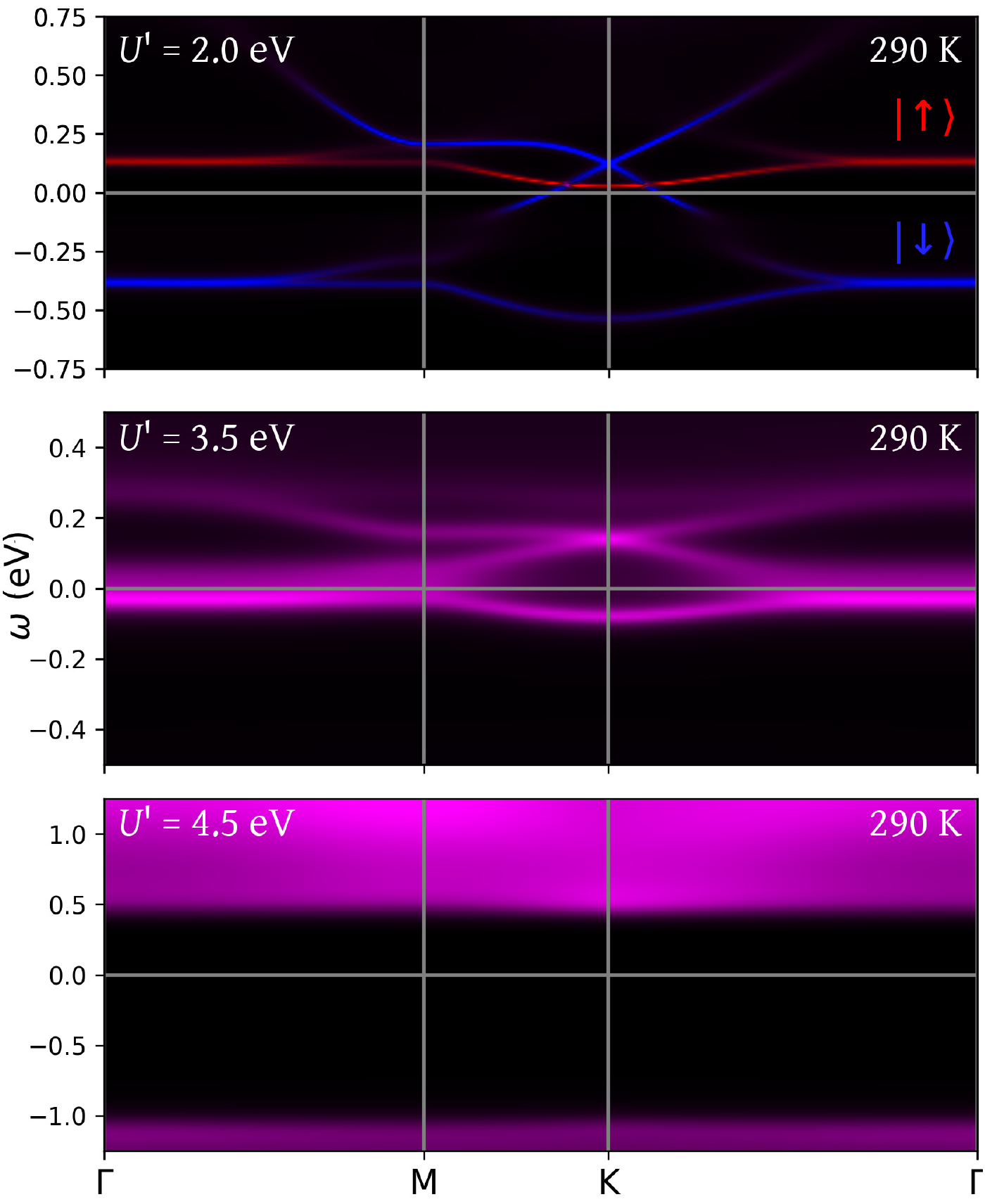}
\caption{\label{fig:awk}(Color online) $k$-resolved spectral functions
$[$Eq.~(\ref{eq:Awk}$)]$ of the uniform structure calculated with DMFT at room
temperature ($T$\,=\,290\,K) for the FM phase ($U'$\,=\,2\,eV), the PM phase
($U'$\,=\,3.5\,eV) and the AOI phase ($U'$\,=\,4.5\,eV). In the FM phase, the
Dirac point at K lies $\sim$100\,meV above the Fermi level.   The two colors
correspond to the spectral weight in the majority ($|\!\downarrow\rangle$) or
minority ($|\!\uparrow\rangle$) channel.  }
\end{figure}

We start with the uniform structure, for which DFT+$U$ calculations yield a FM
Dirac metal, nearly independent of the $U$ value.  Our DMFT ($U'$,$T$) phase
diagram (Fig.~\ref{fig:phasediag}, left) reveals a much more involved picture,
with four different phases: a ferromagnetic metal (FM) at low $U'$, a
paramagnetic metal (PM), an antiferro-orbitally ordered insulator (AOI), and a
paramagnetic insulator (PI); see Fig.~\ref{fig:aw} and Fig.~\ref{fig:awk} for
the $k$-integrated and $k$-resolved $e_g$ spectral functions, respectively.
The long-range ferromagnetic ordering transition temperature $T_{\text{C}}$
depends on the onsite Coulomb repulsion: while $U'$\,$\leq$\,3\,eV yield a
ferromagnetic state at room temperature, larger $U'$\,$\geq$\,3.5\,eV strongly
disfavor spin polarization in the studied temperature range.   In contrast, the
metal-insulator transition (the thick line in Fig.~\ref{fig:phasediag}) occurs
at the critical $U'$ which is slightly smaller than 4\,eV.  The nearly vertical
line separating PM and insulating phases indicates thermal fluctuations play a
minor role in the metal-insulator transition.  In the insulating part of the
phase diagram, the high-temperature paramagnetic phase (PI) develops an orbital
polarization upon cooling, with a gradual crossover to the AOI phase.

\begin{figure}[tb]
\includegraphics[width=8.6cm]{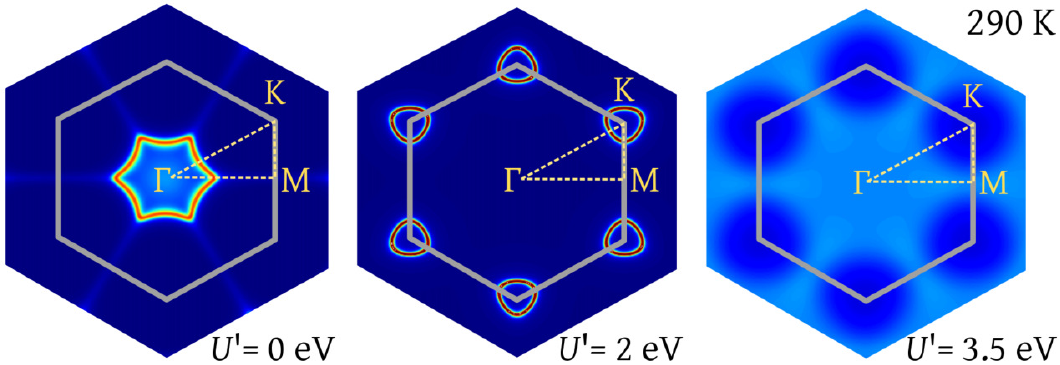}
\caption{\label{fig:fs}(Color online)  Fermi surfaces of the uniform structure
at room temperature.  Left: the non-interacting Hamiltonian (DFT).  Central:
the ferromagnetic metal ($U'$\,=\,2\,eV).  Right: the paramagnetic metal
($U'$\,=\,3.5\,eV).
}
\end{figure}

\paragraph{FM phase}  The existence of a FM phase is seemingly in agreement
with the DFT+$U$ results that yield a Dirac metal state for the uniform
structure.  However, the spin-resolved spectral function $A(\omega,k)$
(Fig.~\ref{fig:awk}, top) reveals that our FM phase is not a Dirac metal:  the
band crossing at the K point lies $\sim$0.1\,eV above the Fermi level
(Fig.~\ref{fig:awk}).  This effect comes primarily from the real part of the
self-energy, which shifts the majority states towards higher frequencies.
Instead, the majority states at the Fermi surface form a loop around the K
point (Fig.~\ref{fig:fs}, middle).

\paragraph{PM phase} Above $T_{\text{C}}$, the LNO bilayer is a paramagnetic
metal, where both Ni sites, both orbitals and both spin channels are equally
occupied with 0.25 electrons per site and orbital.  Correlation effects
manifest themselves in the sizable quasiparticle renormalization
$[$Eq.~(\ref{eq:Z})$]$ which amounts to $\sim$0.35--0.60 depending on the $U'$
value.  The spectral function (Fig.~\ref{fig:awk}, middle) shows a weakly
dispersive feature at the Fermi level, which is largely broadened by the
enhanced $\text{Im}\Sigma(0)$.  As a result, the Fermi surface plot lacks 
sharp features (Fig.~\ref{fig:fs}, right).

\paragraph{PI and AOI phases} Similar to the PM phase, both orbitals and both
spin channels are equally populated also in the PI phase, but the spectral
function has a gap which grows with $U'$.  There is an orbital
disproportionation between the two $e_g$ orbitals setting in at $\sim$350\,K,
and already at room temperature, a sizable orbital polarization develops.  The
two neighboring NiO$_6$ octahedra have different predominantly occupied
orbitals, giving rise to an antiferro-orbital order (AOI).  Interestingly, this
spontaneous symmetry breaking occurs despite the degeneracy of the $e_g$
orbitals, and hence is of a purely electronic origin.  The spectral function
(Fig.~\ref{fig:awk}, bottom) shows a wide gap between two incoherent continua
--- the lower and the upper Hubbard bands.

\begin{figure}[tb]
\includegraphics[width=8.6cm]{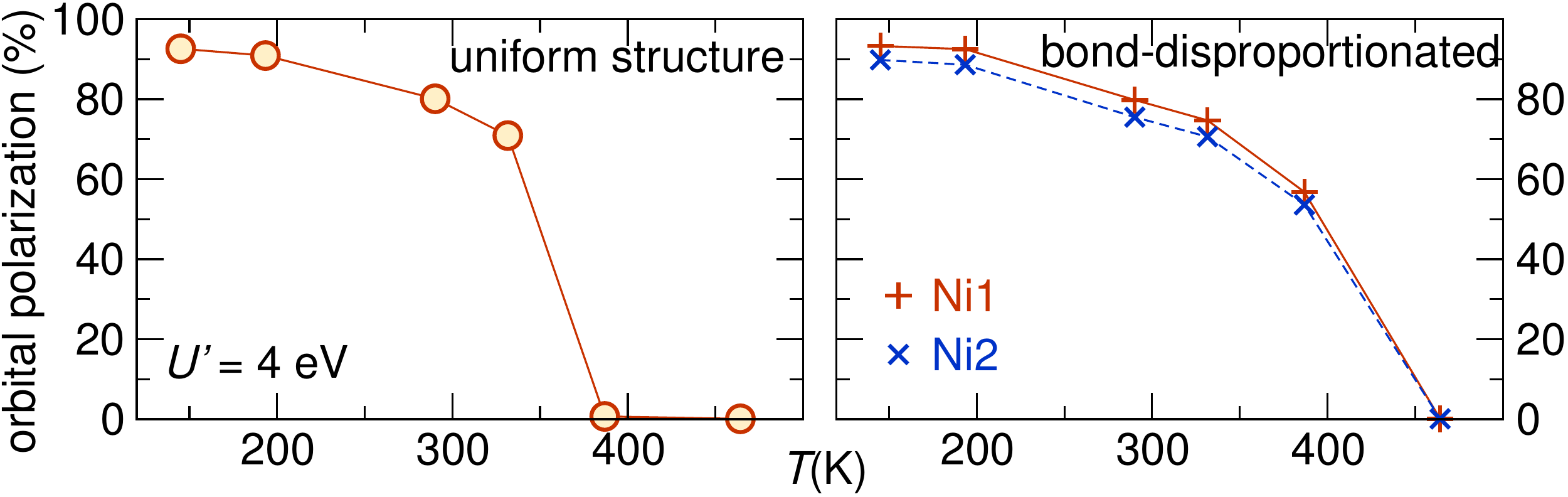}
\caption{\label{fig:orb}(Color online)
DMFT orbital polarization
$p=|(n_{3z^2-r^2}-n_{x^2-y^2})/(n_{3z^2-r^2}+n_{x^2-y^2})|$ in the insulating
phase ($U'$\,=\,4\,eV) as a function of temperature.  The lines are a guide to
the eye.
}
\end{figure}

The degree of the orbital polarization $p$ in nickelates is typically
defined~\cite{wu13, park16} as

\begin{equation}
\label{eq:orb}
p = \left|\frac{n_{3z^2-r^2}-n_{x^2-y^2}}{n_{3z^2-r^2}+n_{x^2-y^2}}\right|,
\end{equation}

where $n_{3z^2-r^2}$ and $n_{x^2-y^2}$ are orbital occupations (a summation
over both spin channels is implied).  The polarization $p$ is shown in
Fig.~\ref{fig:orb} (left) as a function of temperature for $U'$\,=\,4\,eV.  A
sharp increase of orbital polarization is seen below $\sim$350\,K, signaling
the phase transition from the PI to the AOI phase.

\subsection{Bond-disproportionated structure}
For the BD structure, DFT+$U$ calculations with $U$ from a reasonable range
yield a semiconductor with a gap of about 0.05\,eV~\cite{doennig14}, in
contrast to the Dirac metal state of the uniform structure.  Surprisingly, our
DFT+DMFT phase diagram for the BD structure (Fig.~\ref{fig:phasediag}, right)
is very similar to that of the uniform structure.  The four emerging phases are
analogous to those of the uniform structure, expect for the slight charge
disproportionation between the Ni sites that naturally occurs for a BD
structure~\cite{han11,park12}.  Two noticeable differences are i) the shift of
boundaries of both phase transitions (FM$\rightarrow$PM and
PM$\rightarrow$PI/AOI) towards larger $U'$ values and ii) the crossover between
the AOI and PI phases showing an even weaker dependence on $U'$ than in the
uniform case.  Please note that the charge disproportionation of the starting
BD Hamiltonian is very small and hardly affected by the DMFT correlations.
Instead, DMFT correlations support again the orbital polarization, see
Fig.~\ref{fig:orb}~(right), which is not present in the DFT-derived BD Wannier
Hamiltonian.

\section{\label{sec:disc}Discussion}
\subsection{\label{sec:topo}Topological properties}
The honeycomb lattice has an excellent potential for the formation of
topological edge states. The emergence of topological states in (111) bilayers
of $e_g$ electrons has been addressed on the model level~\cite{xiao11, ruegg11}
and in the context of nickelate heterostructures~\cite{yang11, ruegg12,
ruegg13, okamoto14}.   Hartree-Fock calculations~\cite{yang11, ruegg11,
ruegg12} yield a rich phase diagram with orbitally ordered and topological
phases, but direct DFT+$U$ calculations favor a conventional
ferromagnetic phase~\cite{ruegg12}.  Lattice distortions, and in particular,
the breathing distortion can drive the system away from a topological
phase~\cite{ruegg13}, as confirmed by direct DFT+$U$ calculations for LNO
bilayers~\cite{doennig14}.  But in the absence of lattice distortions, DFT+$U$
yields a Dirac metal state.

\begin{figure*}[t]
\includegraphics[width=\textwidth]{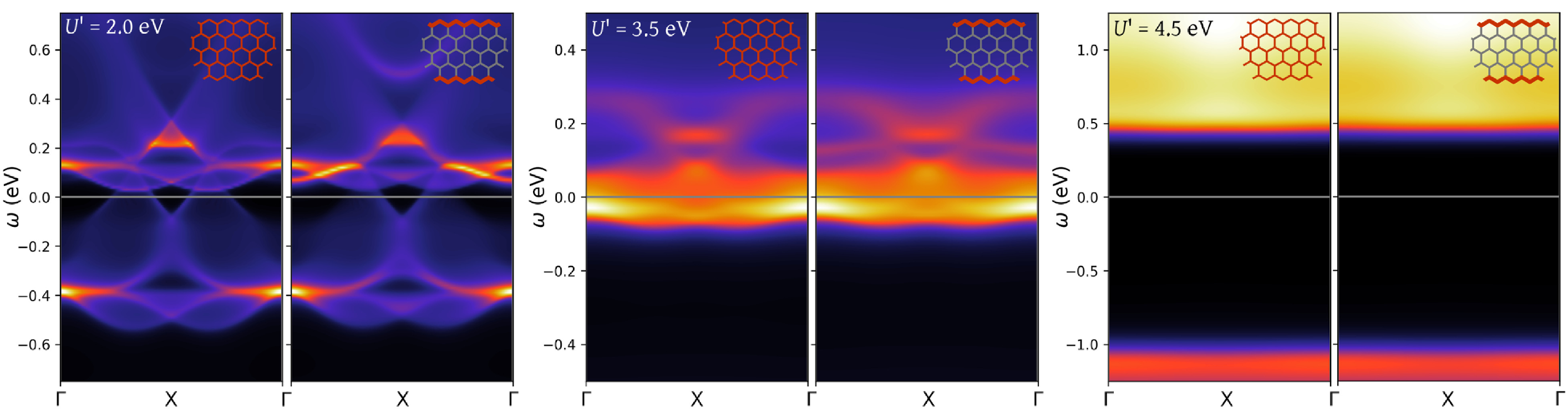}
\caption{\label{fig:awkx}(Color online) DFT+DMFT $A(k_x,\omega)$ at room
temperature on cylinders of 50 unit cells.   In each panel, the left spectrum
shows the total $A(k_x,\omega)$ $[$Eq.~(\ref{eq:Awkx})$]$, while the right
spectrum shows the weight of the edge states $A^{\text{edge}}(k_x,\omega)$
$[$Eq.~(\ref{eq:Awkx_edge})$]$, as schematically depicted in the top-right
corners of each plot. Note the topological edge state visible in the
edge-projected (right) spectrum lying 0.1-0.2\,eV above the Fermi energy for
the $U'$\,=\,2.0\,eV (left panel).}
\end{figure*}

In the above studies, with the notable exception of Ref.~\onlinecite{okamoto14},
electronic correlations were either neglected or taken into account at the
Hartree-Fock level.  DMFT accounts for all local Feynman diagrams, and in this
way represents a systematic and substantial improvement over the Hartree-Fock
method.   On the DMFT level, the electronic correlations are described by the
frequency-dependent self-energy.  In general, the self-energy is a matrix with
nonzero orbital offdiagonal elements.  However, for LNO bilayers, the proximity
of the Ni--O--Ni angles (165.81 and 165.58$^{\circ}$ in the uniform and the BD
structure, respectively) to 180$^{\circ}$ leads to vanishingly small
offdiagonal elements between $3z^2-r^2$ and $x^2-y^2$ orbitals.  As a result,
we can safely neglect offdiagonal elements of the hybridization function
$F(i\omega_n)$ in our impurity problems, leading to the self-energies
$\Sigma(i\omega_n)$ that are diagonal in the site-orbital-spin basis.

The resulting DMFT self-energies are used to calculate the interacting Greens
function using Eq.~(\ref{eq:Gwedge}) and subsequently, the spectral functions
using Eqs.~(\ref{eq:Awkx}) and (\ref{eq:Awkx_edge}).  
To this end, we use the DMFT self-energy of our bulk calculation for all sites
and consider periodic boundary conditions along the $x$ axis and open boundary
conditions for the $y$ axis, leading to the cylinder geometry.  The spectral
functions for the FM, PM, and AOI states are shown in Fig.~\ref{fig:awkx}.
Only the FM state shows a distinct edge state, which however lies entirely in
the unoccupied part of the spectrum, $~\sim$0.1--0.2\,eV above the Fermi
energy.  This agrees with the position of the Dirac point in
Fig.~\ref{fig:awk}~(top). Both PM and AOI phases yield very incoherent
features, without any distinct edge states.  We therefore conclude that the
emergence of topological states in (111) LNO bilayers is unlikely.

\subsection{\label{sec:compar_model}Comparison with model DMFT calculations} By
performing DFT+DMFT calculations in the $d$-basis, we actually solve a two-site
two-orbital Hubbard model at quarter filling ($n=1$).  The phase diagram of
this model based on a simplified (typically, semi-elliptic) density of states
has been studied in the literature, in particular within DMFT, and it is
tempting to put our DFT+DMFT results into this context.

The early Hirsch-Fye QMC results showed a remarkable stabilization of
ferromagnetism in a two-orbital Hubbard model away from half-filling due to the
Hund's coupling~\cite{held98}. A pronounced tendency towards orbital ordering
for the quarter filling has been found in the same study~\cite{held98}. The
interplay between magnetic and orbital ordering has been further addressed by
Peters~\emph{et~al.}~\cite{peters10a,peters10b}, who employed numerical
renormalization group (NRG) to solve auxiliary impurity problems.  For
quarter-filling at zero temperature, they found a first-order metal-insulator
transition between the FM state and the ferromagnetic counterpart of our AOI
state, driven by the increased interorbital interaction $U'$~\cite{peters10a}.
Looking at our DFT+DMFT phase diagrams, one may speculate that the FM metallic
phase extends to large $U'$ values at lower temperatures than accessible by
CT-QMC and shown in Fig.~\ref{fig:phasediag}.  Further, the AOI order actually
supports an insulating ferromagnetic phase through superexchange. Hence one may
speculate further that at lower temperatures, an additional ferromagnetic ordering
occurs in the AOI phase. This would yield two close-by ferromagnetic phases,
similar as was discussed in Ref.~\onlinecite{peters10a} for a
two-orbital model.  In this case, a huge magnetoresistance can be
expected~\cite{peters10a}.

A subsequent paper reports a detailed model study of the quarter-filled
case~\cite{peters10b}. The ground state phase diagram features, in addition to
the two ferromagnetic and a paramagnetic phase, also the AOI phase, albeit in a
narrow region of the phase diagram, where the inter-orbital repulsion $U'$
largely exceeds the Hund's exchange.  One of the main results of
Ref.~\onlinecite{peters10b} is the stabilization of the orbital order without
Jahn-Teller distortions. Our DFT+DMFT calculations not only corroborate this
conclusion, but strengthen it further: the presence of the AOI phase in the BD
structure implies that the antiferro-orbital order is resilient to the
competing mechanism of a breathing distortion.

\subsection{\label{sec:compar_exp}Comparison with experiments}
Recent transport measurements on 2LNO/4LAO heterostructures yield a band gap of
120\,meV~\cite{wei16}, which grows if the thickness of the LAO layer increases,
which has been argued to stem from the accumulation of defects in thicker
layers.  Nevertheless, the insulating nature of the LNO bilayers can be
regarded as a sound experimental result as it also concurs with the earlier
report~\cite{middey12}.  Thus, we infer that the inter-orbital repulsion $U'$
in LNO bilayers exceeds 4\,eV, as only such high values yield an insulating
phase (Fig.~\ref{fig:phasediag}).

At the first glance, we do not have arguments in favor or against the breathing
distortion:  the insulating part of the phase diagram is similar for both
structures, and the DFT+$U$ energies are essentially degenerate. But the bare
emergence of the AOI phase already indicates a tendency towards a cooperative
Jahn-Teller distortion (orbital ordering).  Hence, in the insulating state, the
electronic degrees of freedom disfavor the competing mechanism of a breathing
distortion even in the BD structure.  The presence of such an orbital order can
be verified experimentally by measuring x-ray linear dichroism (see
Ref.~\onlinecite{disa15}), because both the uniform and the BD structure show a very
strong orbital polarization $p$ already at room temperature
(Fig.~\ref{fig:orb}).  Nickelate heterostructures with a sizable orbital
polarization do exist for the (001) case~\cite{disa15}.  In contrast to the
(001) case, the AOI state in (111) LNO bilayers is an insulating state, and a
large orbital polarization is easier to achieve.  In several cases, the
Mott-Hubbard metal-insulator transition is indeed accompanied by an orbital
polarization, e.g.\ in V$_2$O$_3$~\cite{keller04} and SrVO$_3$
films~\cite{zhong14}. 

Magnetic properties of LNO bilayers remain hitherto unexplored, but the recent
study of NdNiO$_3$ bilayers on LAO, reporting antiferromagnetic correlations
and orbital order~\cite{middey16}, demonstrates that an experimental insight is
feasible.  According to our DFT+DMFT results, the insulating phases, PI and
AOI, do not show any magnetic order above $\sim$150\,K, and it would be
interesting to verify this result experimentally.  

A good agreement between our DFT+DMFT and model DMFT
studies~(Sec.~\ref{sec:compar_model}) gives hope that the low-temperature
physics of (111) LNO bilayers can be even more exciting. In particular, a very
high magnetoresistance was found at the boundary between the
antiferro-orbitally-ordered (insulating) and the orbitally-disordered
(metallic) ferromagnetic phases in the model DMFT~\cite{peters10b}. Although we
do not see the former phase in the phase diagram (Fig.~\ref{fig:phasediag}), it
may become stabilized by cooling. Unfortunately, performing CT-QMC at low
temperature becomes prohibitively challenging, although very recent
developments such as the superstate sampling method~\cite{kowalski18} can
largely alleviate the computational effort.  Nonetheless our phase diagram sets
the stage of what orders can be expected in experiment.

\section{\label{sec:summary}Summary and outlook}
Using DFT+DMFT calculations, we evaluated the phase diagram of a (111) oxide
heterostructure formed by LaNiO$_3$ bilayers interleaved with four layers of
LaAlO$_3$, in a wide range of temperatures and the values of the inter-orbital
Coulomb repulsion $U'$.  Independent of the presence or absence of breathing
distortions that are typical for bulk nickelates, we find four phases:  a
ferromagnetic and a paramagnetic metal, a paramagnetic insulator, as well as an
antiferro-orbitally-ordered insulator.  Spectral functions calculated on
cylinders feature edge states in the ferromagnetic metallic state, whereas both
insulating phases are topologically trivial.  Taking the experimentally
observed activated behavior as an indication for a insulating state, we argue
that LaNiO$_3$ bilayers can develop sizable orbital polarization at room
temperature.  Based on earlier model DMFT studies, we can expect ferromagnetic
ordering at lower temperatures, offering an intriguing possibility of a
transition between a metallic and insulating ferromagnetic phases, with a
concomitant high magnetoresistance.

Compared to DFT+$U$, DFT+DMFT provides a more realistic treatment of electronic
correlations and gives access to finite temperature properties.  However, DMFT
is restricted to local correlations.  For a quasi-2D system
with the low coordination number such as nickelate (111) bilayers, nonlocal
correlation effects can play an important role.  A natural extension of our
study would be the application of cluster~\cite{lichtenstein00,kotliar01} or
diagrammatic~\cite{rohringer18} extensions of DMFT to the phase diagram of
2LNO/4LAO.

\begin{acknowledgments}
We acknowledge financial support by European Research Council under the
European Union's Seventh Framework Program (FP/2007-2013)/ERC through grant
agreement n.\ 306447. OJ was supported by the Austrian Science Fund (FWF)
through the Lise Meitner programme, project no.\ M2050.  We thank Marta Gibert,
Sumanta Bhandary, and Gang Li for fruitful discussions. Calculations have been
done on the Vienna Scientific Cluster~(VSC).
\end{acknowledgments}

%

\onecolumngrid
\newpage
\appendix*

\section{DFT+$U$-optimized crystal structures of 2LNO/4LAO (111) heterostructures}

Crystal structures used for DFT+DMFT calculations are provided in Table~\ref{tab:str}.

\begin{table}[h]
\caption{\label{tab:str}  Atomic coordinates of the uniform and the bond-disproportionated structures used for DFT+DMFT calculations.
Space group $P3$\,(143), $a$\,=\,5.3646\,\r{A}, $c$\,=\,13.30\,\r{A}.}
\begin{ruledtabular}
\begin{tabular}{l c r r r c r r r}
\multirow{2}{*}{atom} & \multirow{2}{*}{Wyckoff position} & \multicolumn{3}{c}{uniform structure} & & \multicolumn{3}{c}{bond disproportionated} \\
& & $x/a$ & $y/b$ & $z/c$ & & $x/a$ & $y/b$ & $z/c$\\ \hline
Ni & $1b$ & $\frac13$ & $\frac23$ & $-0.08557$& & $\frac13$ & $\frac23$ & $-0.08477$\\
Ni & $1c$ & $\frac23$ & $\frac13$ &   0.08557 & & $\frac23$ & $\frac13$ &   0.08678\\
Al & $1a$ &         0 &         0 &   0.25335 & &         0 &         0 &   0.25466\\
Al & $1b$ & $\frac13$ & $\frac23$ &   0.41745 & & $\frac13$ & $\frac23$ &   0.41828\\
Al & $1c$ & $\frac23$ & $\frac13$ &   0.58255 & & $\frac23$ & $\frac13$ &   0.58333\\
Al & $1a$ & 0         &         0 &   0.74665 & &         0 &         0 &   0.74754\\
La & $1a$ & 0         &         0 &   0.00000 & &         0 &         0 &   0.00000\\
La & $1b$ & $\frac13$ & $\frac23$ &   0.16501 & & $\frac13$ & $\frac23$ &   0.16584\\
La & $1c$ & $\frac23$ & $\frac13$ &   0.33234 & & $\frac23$ & $\frac13$ &   0.33393\\
La & $1a$ & 0         &         0 &   0.50000 & &         0 &         0 &   0.50145\\
La & $1b$ & $\frac13$ & $\frac23$ &   0.66766 & & $\frac13$ & $\frac23$ &   0.66908\\
La & $1c$ & $\frac23$ & $\frac13$ &   0.83499 & & $\frac23$ & $\frac13$ &   0.83563\\
O  & $3d$ &  0.00001  &   0.45619 &   0.00000 & &   0.45754 &   0.00284 &   0.00222\\
O  & $3d$ &  0.32833  &   0.20833 &   0.16906 & &   0.32947 &   0.20851 &   0.17008\\
O  & $3d$ &  0.12891  &   0.33333 &   0.33440 & &   0.12916 &   0.33333 &   0.33495\\
O  & $3d$ &  0.00000  &   0.53725 &   0.50000 & &   0.00000 &   0.53765 &   0.50040\\
O  & $3d$ &  0.33333  &   0.12892 &   0.66560 & &   0.33333 &   0.12845 &   0.66596\\
O  & $3d$ &  0.20833  &   0.32833 &   0.83094 & &   0.20853 &   0.32728 &   0.83139\\
\end{tabular}
\end{ruledtabular}
\end{table}
\end{document}